\documentstyle[aps,twocolumn,psfig,prl]{revtex}

\def \recent {herzel,bart,barrat,monasson,newman,weigt,Letter,newman2}

\begin{document}

\draft

\title{Infinite Characteristic Length in Small-World Systems}

\author{Cristian F.~Moukarzel and Marcio Argollo de Menezes}

\address{ Instituto de F\'{\i}sica, Universidade ~Federal Fluminense, CEP
  24210-340, Niter\'oi, RJ, Brazil}

\maketitle

\begin{abstract}
  It was recently claimed that on $d$-dimensional small-world networks with a
  density $p$ of shortcuts, the typical separation $s(p) \sim p^{-1/d}$
  between shortcut-ends is a characteristic length for shortest-paths
  (M.~E.~J.~Newman and D.~J.~Watts, ``Scaling and percolation in the
  small-world network model'' {\it cond-mat/9904419}). This contradicts an
  earlier argument suggesting that no finite characteristic length can be
  defined for bilocal observables on these systems (M.~Argollo de Menezes,
  C.~Moukarzel and T.~J.~P.~Penna, ``First-order transition in small-world
  networks'', {\it cont-mat/9903426}). We show analytically, and confirm by
  numerical simulation, that shortest-path lengths $\ell(r)$ behave as
  $\ell(r) \sim r$ for $r < r_c$ and as $\ell(r) \sim r_c$ for $r>r_c$, where
  $r$ is the Euclidean separation between two points and $r_c(p,L) \sim
  p^{-1/d} \log(L^d p)$.  This shows that the mean separation $s$ between
  shortcut-ends is \emph{not} a relevant length-scale for shortest-paths. The
  true characteristic length $r_c(p,L)$ \emph{diverges} with system size $L$
  no matter the value of $p$. Therefore no finite characteristic length can be
  defined for small-world networks in the thermodynamic limit.
\end{abstract}


The recent observation~\cite{watts} that a non-extensive amount of long-range
bonds, or ``shortcuts'', on a regular $d$-dimensional lattice suffices to
drastically shorten shortest-path distances, has triggered appreciable
interest in these so-called ``small-world'' networks\cite{\recent}.

Small-world networks consist of a regular $d$-dimensional lattice of linear
size $L$ to which $L^d p$ long-ranged bonds have been added, connecting
randomly chosen pairs of sites. Several recent studies have concentrated on
the behavior of the average shortest-path distance
$\overline{\ell}=1/N^2\sum_{<ij>} <\ell_{ij}>$, where $\ell_{ij}$ is the
minimum number of links that must be traversed to join sites $i$ and $j$, and
$<>$ means average over disorder realizations. On a regular lattice one has
$\overline{\ell} \sim L$, while on a random graph of $L^d$ sites,
$\overline{\ell} \sim \log(L)$.  On small-world networks, which interpolate
between these two limits, one finds~\cite{watts,bart,newman} $\overline{\ell}
\sim L$ if $L^d p <<1$, but $\overline{\ell} \sim \log{L}$ if $L^d p >>1$.
Thus on a large system a small density of shortcuts is enough to have
shortest-path distances which are characteristic of random graphs.

It has been suggested~\cite{bart} that for any fixed density $p$ of
long-ranged bonds, a \emph{crossover-size} $L^*(p)$ exists, above which the
average shortest-path distance $\overline{\ell}$ increases only logarithmically with
$L$.  Renormalization Group arguments~\cite{newman} as well as numerical
measurements~\cite{barrat,newman,Letter} indicate that this crossover length
diverges as $p \to 0$ as $L^*(p) \sim p^{-1/d}$.  While some authors see this
result as evidence of a continuous phase transition at
$p=0$~\cite{newman,newman2}, it has also been argued~\cite{Letter} that no
finite correlation length can be defined on these systems, and thus that the
small-world transition is first-order.
  
Clearly, the normalized shortest-path distance ${{\mathcal{L}}}(p)=
\overline{\ell}/L$ undergoes a discontinuity at $p=0$ in the thermodynamic
limit. Thus, the ``small-world'' transition is technically a
\emph{first-order}, or discontinuous, transition~\cite{Letter}. A subtler and
still controversial point~\cite{newman,Letter,newman2} is whether it is
possible to identify a correlation length $\xi(p)$ diverging at $p=0$, i.e. if
the small-world transition is associated with some sort of \emph{critical
  behavior}. This would mean that $p=0$ is a \emph{first-order critical
  point}(FOCP)~\cite{FB}. As discussed by Fisher and Berker~\cite{FB}, a FOCP
is a critical point where the thermal eigen-exponent attains its maximum
possible value $y_t=d$, thus allowing the \emph{coexistence} of two different
critical phases.  Consequently at a FOCP one has $\nu = 1/y_t = 1/d$, and some
finite-size corrections will be dictated by this exponent.

A FOCP appears for example in one-dimensional percolation~\cite{Stauffer},
where the order parameter $P_{\infty}$ (the density of the spanning cluster)
is discontinuous at $p_c=1$, while the correlation length $\xi$ diverges as $p
\to 1$ as $\xi \sim (1-p)^{-1}$. Finite-size effects as e.g. the width of the
critical region are of order $L^{-1}$. Because of this, $P_{\infty}$ is
non-zero even for $p\neq 1$, if $L < \xi \sim (1-p)^{-1}$.  Another example of
a FOCP is the Ising model in one dimension, where the magnetization density
$<m>(T)$ is discontinuous at $T=0$ while the correlation length $\xi$ diverges
as $\xi \sim e^{J/T}$.

At a normal first-order point, on the other hand, no critical behavior exists
(i.e.  all relevant $L$-independent length-scales remain finite), but the
eigen-exponent $d$ also shows up in finite-size corrections. In order to
illustrate this, let us consider~\cite{FB} a field-driven first-order
transition at $h=0$ (like e.g.  in the two-dimensional Ising model below
$T_c$). Assume that the central spin in a system of linear size $L$ is pinned
in the direction opposite to the external field $h$, and let us take for
simplicity $T \sim 0$.  If $L$ is large enough, the bulk energy associated
with $h$ will be predominant, and $m(L,h)$ will have the same sign as $h$. On
the other hand if $L$ is small, it will be energetically favorable not to
break any bonds and thus all spins will point in the $-h$ direction.  The
crossover length $L^*$ at which magnetization inversion happens can be
estimated by equating the total field contribution $L^dh$ to the energy cost
associated with the breaking a finite number of bonds around the pinned spin,
and one obtains $L^* \sim h^{-1/d}$~\cite{FB}.  So we see that there may be a
diverging crossover size in a first-order transition, although there is no
diverging correlation length in this case.

Thus the fact that finite-size corrections in the \emph{global} observable
$\overline{\ell}$ of small-world networks behave as $L^{-d}$ is compatible both
with a first-order transition and with a first-order critical point at $p=0$.
Clearly it is still necessary to prove or disprove the existence of critical
behavior in small-world networks, and for this purpose we suggest to look at a
bilocal observable (from which the analogous of a correlation function $g(r)$
could be eventually defined), instead of the global observable
$\overline{\ell}$. We study in this work the behavior of $\ell(r)$, the average
shortest-path distance between two points separated by an Euclidean distance
$r$.

%%%%%%%%%%%%%%%%%%%%%%%%%%%%%%%%%%%%%%%%%%%%%%%%%%%%%%%%%%%%%%%%%%%%%%
\begin{figure}[thpb]
\centerline{\psfig{figure=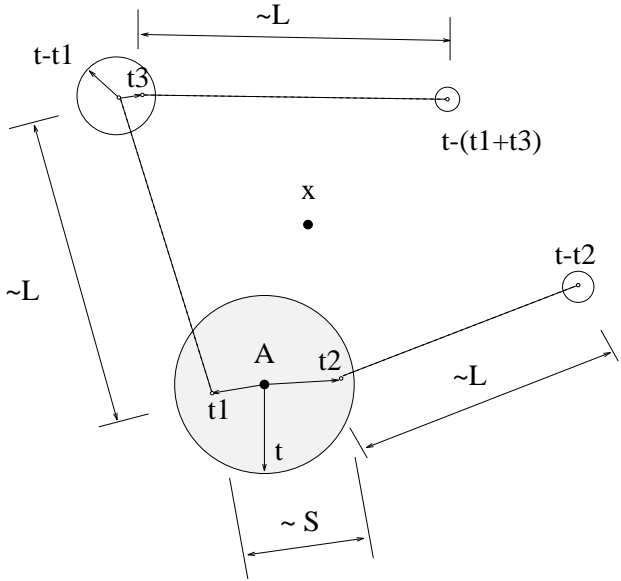,width=9cm,angle=270}}
\caption{ {} We consider a $d$-dimensional continuum of linear size $L$, on
  which a number $pL^d$ of infinitely conducting \emph{shortcuts} (dashed
  lines) have been connected between randomly chosen points. The typical
  separation between nearby shortcut-ends is thus $s\sim p^{-1/d}$. A disease
  starts to spread at $t=0$ from $A$ with unit velocity, and for short times
  only the \emph{primary sphere} (shaded) of radius $t$ becomes infected.
  Each time a sphere hits a shortcut-end, a secondary sphere starts to grow
  from the other end of the shortcut, which is typically located at a distance
  $L$ from there.}
\label{fig:1}
\end{figure}
%%%%%%%%%%%%%%%%%%%%%%%%%%%%%%%%%%%%%%%%%%%%%%%%%%%%%%%%%%%%%%%%%%%%%%

The basic question we wish to answer is whether it is possible to identify a
characteristic length $\xi(p)$ on small-world networks, such that: \emph{i)}
$\xi(p)$ is a length scale dictating the behavior of $\ell(r)$, \emph{ii)}
$\xi(p)$ only depends on $p$ and not on system size $L$, when $L\to \infty$,
and, \emph{iii)} $\xi$ diverges at the critical point $p_c$.

Argollo et al~\cite{Letter} have recently given a simple asymptotic argument
to suggest that, when $L\to \infty$, no \emph{finite characteristic length}
can be defined for bilocal observables such as shortest-path lengths on
small-world networks. Newman and Watts~\cite{newman2} on the other hand,
recently claimed that the characteristic length for shortest-paths is the mean
separation $s(p)\sim p^{-1/d}$ between shortcut-ends.  This quantity clearly
satisfies criteria \emph{ii)} and \emph{iii)} above, but we will show that it
does not satisfy \emph{i)}, i.e. it is \emph{not} relevant for shortest-path
lengths on large systems.

In order to test the ideas above, consider the spread of a disease in a
$d$-dimensional small-world system of linear size $L$, as depicted in
Fig.~\ref{fig:1}. For simplicity we work on the continuum. Assume that, at
$t=0$, an infectious disease starts to spread radially with constant velocity
$v=1$ from A. Let the sphere of radius $t$ grown from A be called ``primary
sphere'' of infected sites, or ``individuals''.  Because of the existence of
shortcuts, which we assume are traversed in zero time, other (secondary)
sources of infection will appear at times $t_k$ at \emph{random locations} in
the available non-infected space.  These times $t_k$ are the times at which an
infection sphere (primary or not) hits one end of a shortcut. We call the
spheres born at the other end of these shortcuts, ``secondary spheres''.  In
this setting, the shortest-path distance $\ell(A,x)$ from $A$ to $x$ is
exactly the \emph{time} $t$ at which a point $x$ is first infected.  If a
point at $x$ is hit by the primary sphere first (before being hit by a
secondary sphere), this will necessarily happen at time $t=d_E(A,x)$, where
$d(A,x)$ is the Euclidean distance from $A$ to $x$. In this case the
shortest-path distance $\ell(A,x)$ is simply $d_E(A,x)$. If on the other hand
a secondary sphere hits $x$ at an \emph{earlier} time $t'$ then
$\ell(A,B)=t'<d_E(A,x)$.

In order to better uncover the physical meaning of a characteristic length
$r_c$ for shortest-path distances $\ell(r)$, let us consider a slightly
modified disease-spreading problem.  Let us assume that the primary version of
the disease (the one that starts at $A$) is lethal, while secondary versions
are modified when traversing shortcuts, and become of a milder, non-lethal
form, which protects individuals who are infected by it against the lethal
form. Therefore noinfected individuals who are hit by the primary sphere will
be killed, while those being hit first by a secondary sphere become immune.
Now assume that a characteristic length $r_c$ exists such that $\ell(r)
\approx r$ for $r<r_c$ and $\ell(r) << r$ for $r>r_c$.  Because of the
relationship between disease-spreading and shortest-path distances described
above, one would conclude that \emph{a)} a person at $x$ is killed by the
disease with a high probability if $d_E(A,x)<r_c$, and \emph{b)} a person at
$x$ is relatively safe if $d_E(A,x)> r_c$. Therefore $r_c$, the characteristic
length of shortest-path distances $\ell(r)$, plays the role of a ``safety
distance'' from the primary infection site $A$ in this modified
disease-spreading model.

Newman and Watts recently provided a clever recursive equation for the total
volume $V(t)$ of infected sites as a function of time~\cite{newman2}, and
solved it exactly in one dimension.  Their result reads
\begin{equation}
V(t) = \frac{1}{2} s \left( e^{4t/s}-1 \right).
\label{eq:volume}
\end{equation}
where $s \sim p^{-1}$ is the average separation between ends of different
shortcuts. Notice that $V(t) \sim t$ for $t<<s$ (only the primary sphere is
important) while later $V(t) \sim e^{t/s}$ when $t>s$ (proliferation of
secondary spheres). Because of this change in the behavior of $V(t)$ at
$t\approx s$, Newman and Watts~\cite{newman2} suggest that $s$ is the
equivalent of a correlation length.

The total infected volume $V(t)$ in (\ref{eq:volume}) is the sum of the
primary volume $V_1(t)$ plus the secondary volume $V_2(t)$. The primary volume
is simply $V_1(t)= t^d\Omega_d$, where $\Omega_d$ is the volume of a
$d$-dimensional hypershpere of unit radius. According to our discussion above,
we are interested in calculating the probability $P_S(r)$ for an individual
located at a distance $r$ from $A$ to become infected by a secondary sphere
before being hit by the primary sphere. Because the secondary volume $V_2(t)$
is \emph{randomly distributed} in the total available space $L^d$, it is easy
to calculate the probability $p_2(t)$ for an individudal to become infected by
the secondary version of the disease at time $t$ \emph{or earlier}. This is
simply the fraction of the total volume that is covered by the secondary
volume: $p_2(t) = V_2(t)/L^d$, and does not depend on the Euclidean distance
from the primary source at $A$. If the individual does not become immunized by
the secondary disease first, it will certainly be killed by the primary
infection at time $t=d_E(A,x)$, when the primary sphere reaches him.  Thus the
relevant quantity is $p_2(t)$ evaluated at $t=d_E(A,x)$, which measures the
probability $P_S(d_E(A,x))$ for an individual at $x$ to survive. In one
dimension one obtains, using (\ref{eq:volume}) and $V_1(t)=2t$,
\begin{equation}
P_S(d_E(A,x)) = \frac{s}{2L}  \left( e^{4d_E(A,x)/s}- \frac{4 d_E(A,x)}{s} - 1 \right),
\label{eq:survival}
\end{equation}
which goes to zero when $L \to \infty$ no matter how large $d_E(A,x)$ is. Thus
there is no \emph{finite} safety distance in the thermodynamic limit, meaning
that $r_c$ diverges with system size.  The size-dependence of the ``safety
distance'' $r_c$ can be estimated from (\ref{eq:survival}) by equating $P_S
\approx$ constant, from which one obtains $r_c \sim s \ln(L/s)$.

These heuristic arguments are confirmed by an exact calculation of
$\ell(r)$~\cite{CFMunpub}, as described in the following.  Newman and Watts
equation for $V(t)$~\cite{newman2} can be exactly solved in all dimensions,
and using this solution it is possible to calculate the average shortest-path
distance $\ell(r)$ between two points separated by an Euclidean distance $r$.
The result is~\cite{CFMunpub}:
\begin{equation}
\ell(r) \sim \left\{
\begin{array}{lll}
r &\hbox{for} &r < r_c = p^{-1/d}\log(p^{1/d}L)
\\
r_c &\hbox{for} &r > r_c 
\end{array}   \right.
\label{eq:ell}
\end{equation}

\begin{figure}[thpb]
  \centerline{\psfig{figure=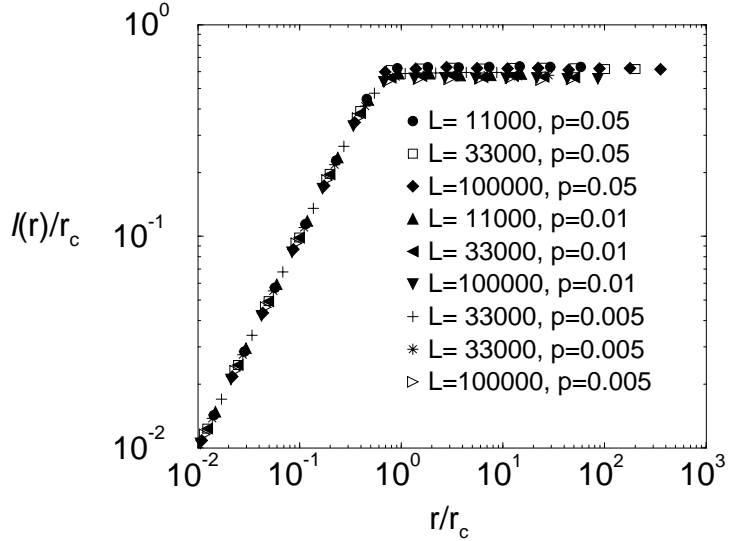,width=9cm,angle=270}}
\caption{ {} Rescaled shortest path distance $\ell(r)/r_c$ versus rescaled
  Euclidean distance $r/r_c$, for one-dimensional small-world networks.  The
  good collapse of these data onto a single curve clearly confirm
  Eq.~(\ref{eq:ell}) in the text. The characteristic length $r_c$ for
  shortest-paths behaves as $p^{-1/d}\log(L^dp)$ and thus diverges, for all
  $p$, with system size $L$.}
\label{fig:2}
\end{figure}

We confirmed this result numerically in one dimension, by measuring the
average shortest path distance $\ell(r)$ as a function of Euclidean separation
$r$, for several values of $L$ and $p$. Equation (\ref{eq:ell}) then means
that the rescaled shortest-path distance $\tilde{\ell}(r) = \ell(r)/r_c$
should be a universal function of the rescaled Euclidean distance
$\tilde{r}=r/r_c$. This is confirmed by the fact that all data displayed in
Fig.~\ref{fig:2} fall on a single curve.

Notice that the characteristic length $r_c\sim s \log(L/s)$ is much larger
than $s$, the mean separation between shortcuts. The reason why $s$ is not a
relevant length scale for shortest-path distances is not difficult to
understand. Consider the shortest-path distance $\ell(A,x)$ between $A$ and
some point $x$ as in Fig.~\ref{fig:1}.  While the average distance one has to
travel from $A$ in order to find a shortcut end is $s$, this shortcut has a
typical length $\sim L$ and thus will take us too far away from $A$ to be
useful in shortening the distance to $x$. 

%%%%%%%%%%%%%%%%%%%%%%%%%%%%%%%%%%%%%%%%%%%%%%%%%%%%%%%%%%%%%%%%%%%%%%
%%%%%%%%%%%%%%%%%%%%%%%%%%%%%%%%%%%%%%%%%%%%%%%%%%%%%%%%%%%%%%%%%%%%%%

According to (\ref{eq:ell}), only when $d_E(A,x)$ is larger than $r_c = s
\log(L^dp)$ is it more ``convenient'' to go from $A$ to $x$ through shortcuts
than directly through the lattice. This result means that the typical shortest
path in the regime of long distances $r >> r_c$ contains
${\mathcal{O}}(\log(L^dp))$ shortcuts, since the typical ``price'' paid for
each shortcut traversed is the average distance between shortcut ends, i.e.
$s$.

Thus $s$ is \emph{not} a relevant scale for shortest-path lengths $\ell(r)$
and therefore it does not play the role of a correlation length as suggested
recently~\cite{newman2}.  A physically more appealing interpretation of $s$ is
to say~\cite{newman2} that it is a diverging \emph{timescale} for the dynamic
process of disease spreading. 

As might be clear by now, the impossibility to define a finite correlation
length is a consequence of the fact that the locations of the secondary
spheres are uncorrelated with the location of the primary infection.  In other
words, while the typical separation between ends of different shortcuts $s(p)$
is finite for $L \to \infty$, the typical separation between both ends of the
same shortcut, which is an important quantity for shortest-paths, scales as
$L$. As a consequence of this, the number of shortcuts with both ends inside a
region of any finite size $\xi$ goes to zero when $L \to \infty$, which
already implies the impossibility to define a finite correlation
length~\cite{Letter}. Thus, although our demonstration here that no finite
correlation length can be defined for shortest-path lengths is more rigorous,
the same conclusion can be reached by using simple arguments~\cite{Letter}.

A different situation would certainly arise if shortcuts had a
length-dependent distribution, for example if each site $i$ is connected with
probability $p$ to another site $j$ chosen with probability $r^{-\alpha}$,
where $r$ is the distance between $i$ and $j$ and $\alpha$ is a free
parameter. For $\alpha \to 0$, this model is the same as discussed here, while
for $\alpha$ large enough one would only have short-range connections and thus
there would be no logarithmic regime in $\overline{\ell}$, even for $p=1$. We
speculate that the transition in $p$ could be shifted to a finite $p_c$ for
some intermediate $\alpha$ values.  We are presently working on this extended
model~\cite{modified}.

\acknowledgments We acknowledge useful discussions with P.~M.~C. de Oliveira,
T.~J.~P.~Penna, D.~J.~Watts, and M.~Newman. This work is supported by CAPES
and FAPERJ.

\end{document}